\documentclass[12pt]{iopart}
\usepackage{iopams}
\usepackage{epsfig}   
\usepackage{graphics}
\newcommand{\beq}{\begin{equation}}
\newcommand{\eeq}{\end{equation}}
\newcommand{\bea}{\begin{eqnarray}}
\newcommand{\eea}{\end{eqnarray}} 
\newcommand{\ba}{\begin{array}}
\newcommand{\ea}{\end{array}}

\begin{document}

\title[Testing the DGP model with ESSENCE]{Testing the DGP model with ESSENCE}

\author{Sara Rydbeck$^1$, Malcolm Fairbairn$^{2,3}$ and Ariel Goobar$^1$}

\address{$^1$Department of Physics, Stockholm University,\\
Albanova University Center, S--106 91 Stockholm, Sweden}
\address{$^2$Perimeter Institute for Theoretical Physics, 31 Caroline Street North,\\
Waterloo, Ontario, Canada N2L 2Y5}
\address{$^3$CERN Theory Division, CH-1211, Geneva 23, Switzerland}

\begin{abstract}
We use the recent supernova data set from the ESSENCE collaboration combined with data from the Supernova Legacy Survey and nearby supernovae to test the DGP brane world model and its generalisations.   Combination of this data with a flatness prior and the position of the peak of the CMB disfavours the DGP model slightly.  Inclusion of the baryon acoustic peak from the Sloan Digital Sky Survey increase the tension of the DGP model with the data, although it is not clear how self consistent this procedure would be without a re-analysis of the survey data in the framework of the DGP cosmology.  Generalisations of the DGP model are tested and constraints on relevant parameters obtained.  
\end{abstract}

\eads{\mailto{sararyd@physto.se}, \mailto{malc@cern.ch}, \mailto{ariel@physto.se}}




\section{Introduction}
Most of the energy in the universe seems to be dark, and many different observations suggest that only a minority of this energy can be dark matter, the rest apparently being some kind of energy component with negative pressure.

The simplest explanation for this dark energy component is a
cosmological constant and people have attempted to explain the
smallness of this energy density and hence its relatively recent
dominance using anthropic arguments \cite{weinbergshapiro} which may concur with modern predictions from string theory \cite{landscape}.
Other explanations exist however, including a class of theories known
as quintessence where the observed acceleration of the universe
results from the stress energy of some rolling scalar field which only
comes to dominance recently \cite{ratra,wetterich}.

A different approach is to ask whether General Relativity breaks down
at large distances, in other words to modify the left-hand/curvature
side of the Einstein equations rather than the
right-hand/stress-energy side.  One of the most studied examples
of such a scenario is the DGP brane world model
\cite{dgp,def1,def2}.  This paper aims to compare the predictions of this theory and its generalisations with the latest cosmological data.

In the DGP model, gravity is trapped on a four dimensional brane world
at short distances, but is able to propagate into a higher dimensional space at
large distances.  The Lagrangian for the model is ($\hbar=c=1, M_{Pl}^2=(8\pi G)^{-1}$)
\begin{equation}
S=-\frac{M^3}{2}\int d^5x\sqrt{-g^{(5)}}R^{(5)}-\frac{M_{Pl}^2}{2}\int d^4x\sqrt{-g^{(4)}}R^{(4)}+\int d^4x\sqrt{-g^{(4)}}{\cal L}_M
\label{DGPlagrangian}
\end{equation}
Due to the different mass scales $M$ and $M_{Pl}$, gravity propagates
differently on the brane and in the bulk. The effect of gravitational
leakage into the bulk will only appear at large distances
($r>L=M_{Pl}^2/2M^3$).

The $tt$ Friedman equation in this theory takes the form
\begin{equation}
H^2-\epsilon\frac{H}{L}=\frac{\rho_M}{3M_{Pl}^2}
\label{DGPfriedman}
\end{equation}
where $\epsilon=\pm 1$. It is $\epsilon=+1$ that gives the late-time
accelerating solutions that could explain dark energy.  The theory
also predicts very small but potentially detectable modifications to
the earth-moon distance \cite{moon}.  There are also changes in
structure formation which could give constraints upon the theory \cite{DGPstructure} (see also \cite{Reboucas}).

Recently the self-accelerating branch of this theory has come under
theoretical attack because it seems to possess ghost-like
instabilities \cite{ghost} while one of the original authors has
argued that the calculational regime within which the instabilities are found is not valid \cite{strong}.  In this work, we will put this
issue to one side and proceed to compare the model's predictions with
the latest data.  Even if it turned out that the DGP model was theoretically suspect, it is interesting to see how different expansion histories may or may not be ruled out by the developing suite of data.

To test theories, we would like to make detailed comparisons with astronomical observations.  There are basically two ways of doing this.  As already mentioned, one is to look at the growth of perturbations in these models and see how they compare with perturbations in $\Lambda$CDM and in the observed galaxy correlation function \cite{DGPstructure}.  Another way is to simply look at the space-time geometry of the universe on the largest length and time scales available in order to reconstruct the expansion history of the universe and compare it to the solutions of the modified gravity models.  This latter approach is the subject of this paper.

Studies regarding the comparison between data and the DGP model based upon expansion history are to date contradictory.  The first paper on the subject was presented by two of the current authors \cite{fairbairngoobar} using the 2005 data release from the SuperNova Legacy Survey (SNLS) \cite{astier06}.  In that work it was argued that the SNLS data combined with the position of the baryon acoustic peak in the Sloan Digital Sky Survey were less consistent with a flat universe in the DGP model  than in the $\Lambda$CDM model.  These conclusions were backed up when Maartens and Majoretto performed the same test using the SNLS data, the baryon acoustic peak and the CMB shift parameter \cite{maartens}.

Later, a new "gold" data set of supernovae was released \cite{riess07}.  Analysis of this data and the CMB shift parameter suggested that a flat universe was completely consistent with the DGP model \cite{scoop}, a conclusion the present authors can confirm, also upon addition of the data from the baryon acoustic peak.  It is interesting to note that it has been suggested that there may be an inconsistency between the different parts of this gold data set \cite{inconsistency}.

The two sets of data come from different instruments - the SNLS supernovae are detected using a combination of imaging at the Canada-France-Hawaii Telescope and spectroscopic studies on large ground based instruments, in particular Gemini, VLT and Keck. The Riess et al. 07 gold sample contains supernovae from the Supernova Cosmology Project, SNLS, the High-Z Team, the GOODS transient survey,  and includes 21 new supernovae at extremely high redshift obtained with the Hubble Space Telescope.  In all cases, low redshift supernovae from independent surveys have been included in the data sets.

Furthermore the supernova magnitudes are obtained from the two different sets of supernova data using two different methods, the SALT algorithm which is used by the SNLS group fits supernovae based upon their light curve and their colour \cite{SALT} as does the MLCS method favoured by Riess et al. , the most recent version being called MLCS2k2 \cite{MLCS}.  The parameters in these two algorithms are obtained by training on low redshift supernovae before they are used to obtain the magnitudes of high redshift supernovae.  SALT and MLCS2k2 differ both in the details of the brightness-shape relation correction and in how extinction/colour corrections are applied.

Recently the ESSENCE group has released a set of 60 supernovae at intermediate to high redshifts \cite{essence}.  This data has been combined with the SNLS data and data at low redshift to form a new data set spanning a large redshift range in detail.  The supernovae in this data set have been analysed using both methods - SALT and MLCS.  In this paper, we combine this new data with the CMB and baryon oscillation data in order to see if the DGP model is favoured or disfavoured.

While the model as it stands has one extra space dimension, it is
possible to imagine generalisations of the model with higher
dimensional bulks, although because of the curvature singularity an infinitely thin 4D brane would
create in more than 5 space-time dimensions, the theory would need to
be regularised at the location of the brane \cite{Gregory}.  We will present a parametrisation of these higher
dimensional generalisations as has been done in previous work
\cite{dvaliturner,chung,fairbairngoobar} and we will introduce an extra
degree of freedom which may go some way to modeling the
regularisations.

\section{\label{difdata}Observational status of Expansion History.}
Observations of high-redshift Type Ia supernovae have been used for
over a decade to map the expansion history of the universe
\cite{goo95,perl95,perl97,garnavich98,riess98,perl98,schmidt98,perl99,knop03,sullivan03,tonry03,barris04,riess04,krisciunas05,nobili05,astier06,conley06,riess07,essence}\footnote{In fact, the very first
detection of a high-z Type Ia SN, SN1988U at z=0.31, was done almost twenty years ago 
by \cite{danish89}. However, the data on this SN was scarce. Thus, this object is normally not
included in the compilation of SNeIa distances.}.
These high-z surveys, combined with low redshift SN sample to anchor the Hubble diagram 
\cite{hamuy96,riess99,jha06}, provided the first direct evidence that the universe
expands at an accelerated rate. Cross-cutting techniques involving
the anisotropies in the cosmic microwave radiation (CMB)
\cite{bernardis,melchiorri,balbi,spergel03,spergel06},
 mass density estimates from X-ray observations
of clusters \cite{allen04} and
more recently weak lensing \cite{hoekstra06}, baryon acoustic
oscillations (BAO) \cite{Eisenstein05,tegmark06} and
the integrated Sachs-Wolfe effect \cite{Boughn02,nolta04}
are apparently continuing to confirm our current understanding of the universe, namely it being spatially flat, as predicted by inflation, and the current expansion 
being dominated by dark energy.

The ESSENCE group has analysed supernovae at intermediate redshifts ($0.1<z<0.8$) using both the MLCS2k2 and the SALT methods to obtain the supernova magnitudes from the light curves and spectra.  When using the MLCS2k2, the group adopt what they refer to as the 'glosz' prior to model $A_V$ - the extinction in the $V$ band of the supernova in the host galaxy \cite{essence}.  They have also fitted higher redshift supernovae ($0.11<z<1.1$) from the SNLS survey \cite{astier06} and a set of nearby supernovae ($z<0.11$) again using both the SALT and MLCS2k2 method to obtain magnitudes of supernovae over a large redshift range.

Measurements of the brightness of Type Ia supernovae as a function of 
redshift are sensitive to the cosmological model via the integration over expansion history in the expression for the luminosity distance ($c=1$)
\begin{equation}
 d_L = 
   \frac{1+z}{H_0\sqrt{|\Omega_k|}} {\cal S}
    \left( \sqrt{|\Omega_k|} \int_{0}^{z} { {d\bar{z} \over E(\bar{z})}} \right)   
\label{lumdim}
\end{equation}
where the function ${\cal S}(x)$ is defined as $\sin(x)$ for 
$\Omega_k<0$ ({\em closed Universe}), 
 $\sinh(x)$ for $\Omega_k >0$ ({\em open Universe}) and  
${\cal S}(x) = x$, and the factor
 $\sqrt{|\Omega_k|}$ is removed for the {\em flat Universe}.   The parameter $E(z) = H(z)/H_0$.   
 
 In our analysis we fit the cosmological model to the supernova data using the luminosity distance above but we also fit to the baryon acoustic peak detected in the SDSS Luminous Red Galaxy survey (LRG) of Eisenstein el al (2005) \cite{Eisenstein05,spergel06}, which constrains the following combination of parameters
\begin{equation}
{ \sqrt{\Omega_M} \over E(z_1)^{1 \over 3} }
\left[{1 \over z_1 \sqrt{|\Omega_k|} } 
{\cal S}\left(\sqrt{|\Omega_k|}\int_0^{z_1} {dz \over E(z)} \right) \right]^{2 \over 3} =0.472 \pm 0.017,
\label{bao}
\end{equation}
where $z_1=0.35$ and ${\cal S}$ and $\Omega_k$ are defined
as in Eq.(\ref{lumdim}). The quoted uncertainty corresponds to one
standard deviation, where a Gaussian probability distribution has been
assumed.

There is some debate \cite{scoop, scoopref} as to whether or not one should use the baryon acoustic peak to constrain models of dark energy which behave differently to a cosmological constant, for two reasons.  The first is that the reconstruction from redshift space to co-moving space required to accurately identify the position of the acoustic peak has been done assuming a constant equation of state \cite{Eisenstein05}.  While one would expect the change in the position of the acoustic peak in an alternative dark energy model where the equation of state is a function of redshift to be small, the correction is difficult to quantify without detailed study for each model in question.  Secondly, in modified models of gravity, one would expect structure formation to proceed differently \cite{DGPstructure}.  Although at the first approximation this would not change the physical co-moving size of the acoustic peak feature in the correlation function, such an effect might create systematic distortions in its reconstruction from redshift space.

For these reasons we will constrain the models with and without the SDSS baryon acoustic peak, we leave it to the reader to decide if they want to pay attention to the BAO constraints on parameter space or not.

Finally, a cleaner measure of geometry is the CMB shift parameter
\cite{bet,maartens} (see however \cite{elgaroy}) - the expansion history of the universe has to be
such that the observed position of the CMB peak corresponds to the
physical horizon size at last scattering.  The photons used to measure
this angular size have passed through the integrated geometry created
by the particular dark energy model in question.  The angular size of
the first peak of the CMB as measured by the WMAP 3 year data
constrains the ratio \cite{WMAP3,wangmukherjee}
\begin{equation}
\sqrt{\Omega_M}\int_{0}^{z} { {d\bar{z} \over E(\bar{z})}}=1.70 \pm 0.03
\label{cmb}
\end{equation}
which can then be applied as another cut to parameter space for each model.

Having described the observations that we will compare with the theoretical models, we can move on the expansion predictions for brane world gravity.

\section{\label{DGPsection}Comparison of the DGP model with the data.}

In this section we will compare the predictions of
the DGP model with the latest cosmological expansion history data set described in the previous section.  As we have discussed in the introduction, the DGP model in its simplest form with one extra flat dimension (\ref{DGPlagrangian}) gives rise to a modified Friedman equation as written in equation (\ref{DGPfriedman}).

We have made a distinction between best model fit and parameter determination - when we talk about 'confidence levels' we are referring to the regions within which the $\chi^2$ values change from the minimum $\chi^2$ value by the critical amounts that are usually used ($\chi^2_{min}+2.3$ for $68\%$ etc. ).  If we use the term '1 $\sigma$', '2 $\sigma$' or '3 $\sigma$' we are referring to respectively the $68\%$, $95\%$ and $99\%$ significances obtained by comparing the $\chi^2$ to the number of degrees of freedom.  If the errors were Gaussian {\it and} the model being fitted was a good one then those two measures of statistical significance should be equivalent.  However, since we do not know if both those criteria are exactly fulfilled, it is important that the reader understands the distinction between the different quantities.

\subsection{Mathematical preliminaries}
It will be interesting to also
consider generalisations of the DGP model which might result from
having a higher dimensional bulk.  Although such models have not been
derived explicitly, it is possible to guess at their possible
form and the way that the Friedman equation would be modified as a
function of the number of extra dimensions \cite{dvaliturner}.  Actual realisations of higher dimensional models of this kind have problems which arise when one considers a delta function of stress energy at the position of the brane \cite{Gregory}. This is solved assuming a
non-zero brane thickness, which means that a potential is required to maintain a nearly massive 4D graviton on the brane \cite{porat}.  This potential will distort the spectra of gravitons in the extra dimensions and also therefore the leakage of gravity into the extra dimensions.  It is not known precisely what form the modifications to the Friedman equations would take in such a situation in the cross over from 4D to higher dimensional physics, we assume that such a regularisation would take the generalised form \cite{prive}
\beq
H^2-\frac{1}{L^2(\beta+(LH)^{n-2})}=\frac{\rho_M}{3M_{Pl}^2}
\eeq 
for zero spatial curvature on the brane. Here $L$ corresponds
to the crossover length scale and $n$ is the number of extra dimensions.  The extra term $\beta$ parameterises the regularisation in cases where $n\neq 1$.

Dividing through by $H_0^2$, we have
\beq
E^2(z)-\frac{\Omega_L}{\beta+\sqrt{\Omega_L}^{(2-n)}{E(z)^{(n-2)}}}=\Omega_M(1+z)^3
\label{dgpe}
\eeq 
where $\Omega_L=(H_0L)^{-2}$ and $E(z) = H(z)/H_0$.  It is easy to see that when $\beta$ dominates the denominator\footnote{try saying that five times quickly}, the $\Omega_L$-term will behave
very much like a cosmological constant term.  As usual, we can constrain one of the parameters in terms of the others by setting
$z=0$ in (\ref{dgpe}), giving 
\beq
\beta=\frac{\Omega_L}{1-\Omega_M}-\sqrt{\Omega_L}^{(2-n)}
\label{constraint}
\eeq
which is the equivalent of the equation $\Omega_\Lambda+\Omega_M=1$ in flat $\Lambda$CDM.

\subsection{Models with $\beta=0$, including original DGP model}

The first models that we constrain are those with the regularisation parameter $\beta$ set to zero.  If there are too many parameters in a dark energy model it is usually rather easy to fit the data for some combination of those parameters, so one can assume priors like imposing a flat universe in order to obtain interesting constraints.  Here, because we have the same number of free parameters\footnote{we have theoretical motivation for $n$ being a discrete integer and each $n$ corresponding to a different model} in these $\beta=0$ models as in the case of $\Lambda$CDM, we are able to include curvature and still get interesting constraints.  We do this simply by replacing the Hubble constant squared $H^2$ with $H^2+ka^{-2}$ \cite{fairbairngoobar}.  This gives us
\beq
H^2+\frac{k}{a^2}-L^{-2}\left[\beta+\left(L\sqrt{H^2+\frac{k}{a^2}}\right)^{n-2}\right]^{-1}=\frac{\rho_M}{3M_{Pl}^2}
\label{curvgeneral}
\eeq 
which for the case of the original DGP model leads to
\begin{equation}
E^2(z)=\Omega_K(1+z)^2+\left(\sqrt{\Omega_M(1+z)^3+\Omega_L/4}+\sqrt{\Omega_L/4}\right)^2
\end{equation}
where the normal definition $\Omega_K=-k/(a H_0)^{2}$ has been used.  

The different models were compared with the two sets of data, one of which contained supernovae that had been analysed using the SALT method of determining magnitudes while the other using the MLCS2k2 method.  The data-set analysed by the ESSENCE collaboration with SALT gave rise to anomalously large values of $\chi^2$ seemingly due to a few outlying data points.   Since the data analysed using MLCS gave $\chi^2$ per degree of freedom rather close to one we have more confidence in this data set and have restricted ourselves to using it. For this data set the intrinsic error in the magnitudes is taken to be 0.1 and the peculiar velocity error in the redshift, which has importance for the lowest redshift supernovae, is assumed to be 400 km/s. To demonstrate the effect of different error assumptions we present the resulting fits with and without inclusion of the peculiar velocity error.

\begin{figure}
\begin{tabular}{cc}
\includegraphics[height=6cm,width=8cm]{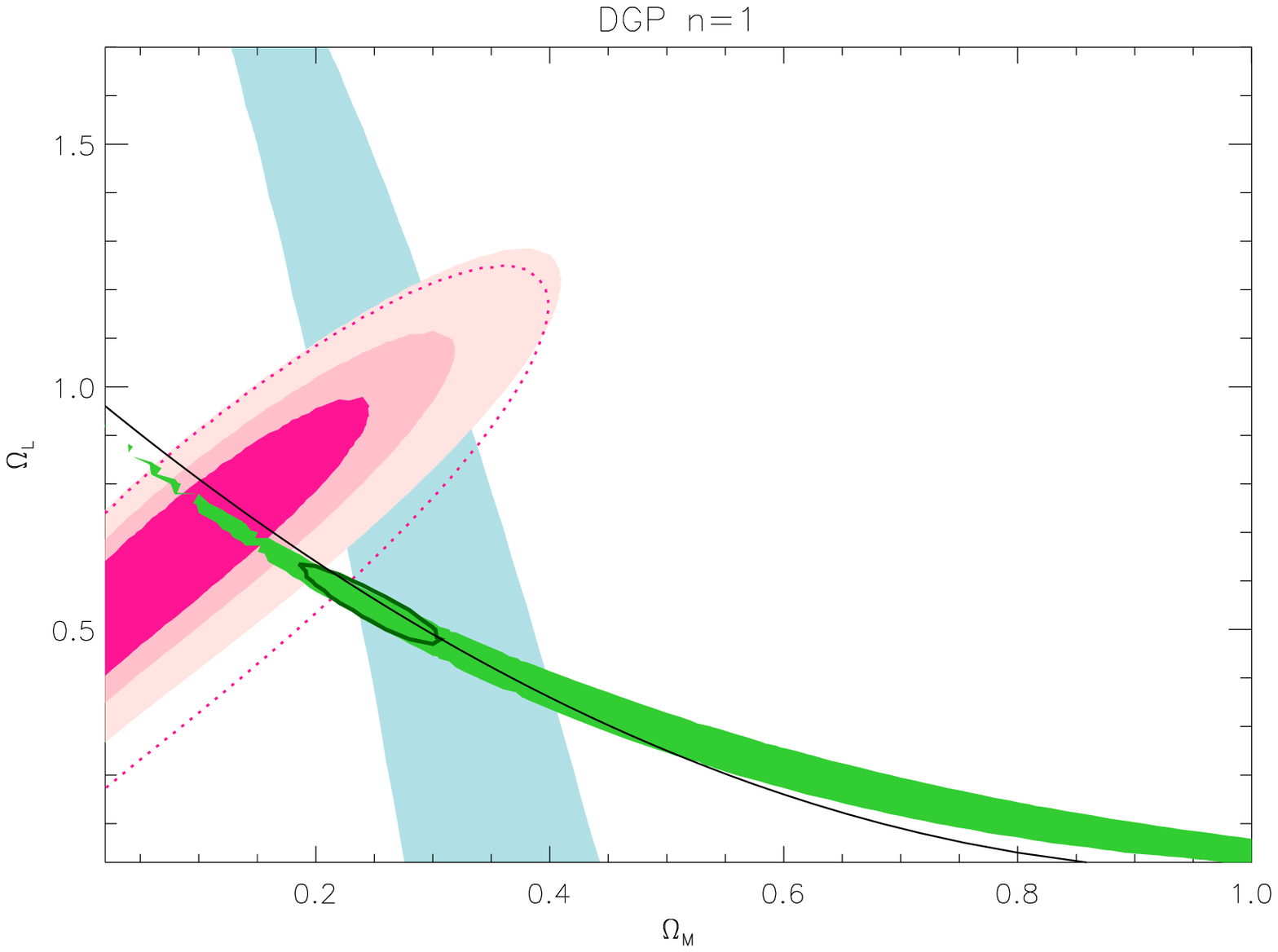}& 
\includegraphics[height=6cm,width=8cm]{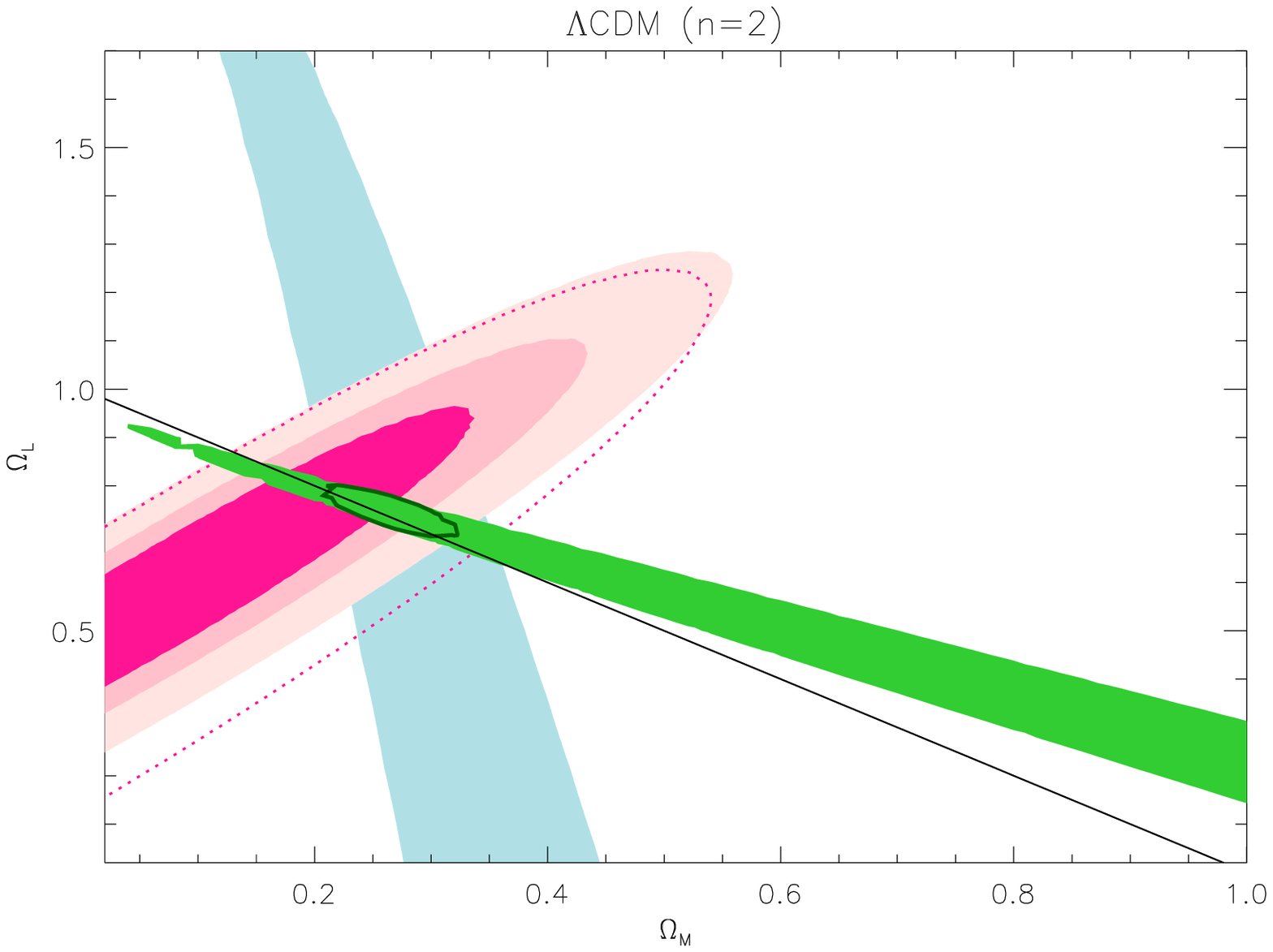}\cr
\includegraphics[height=6cm,width=8cm]{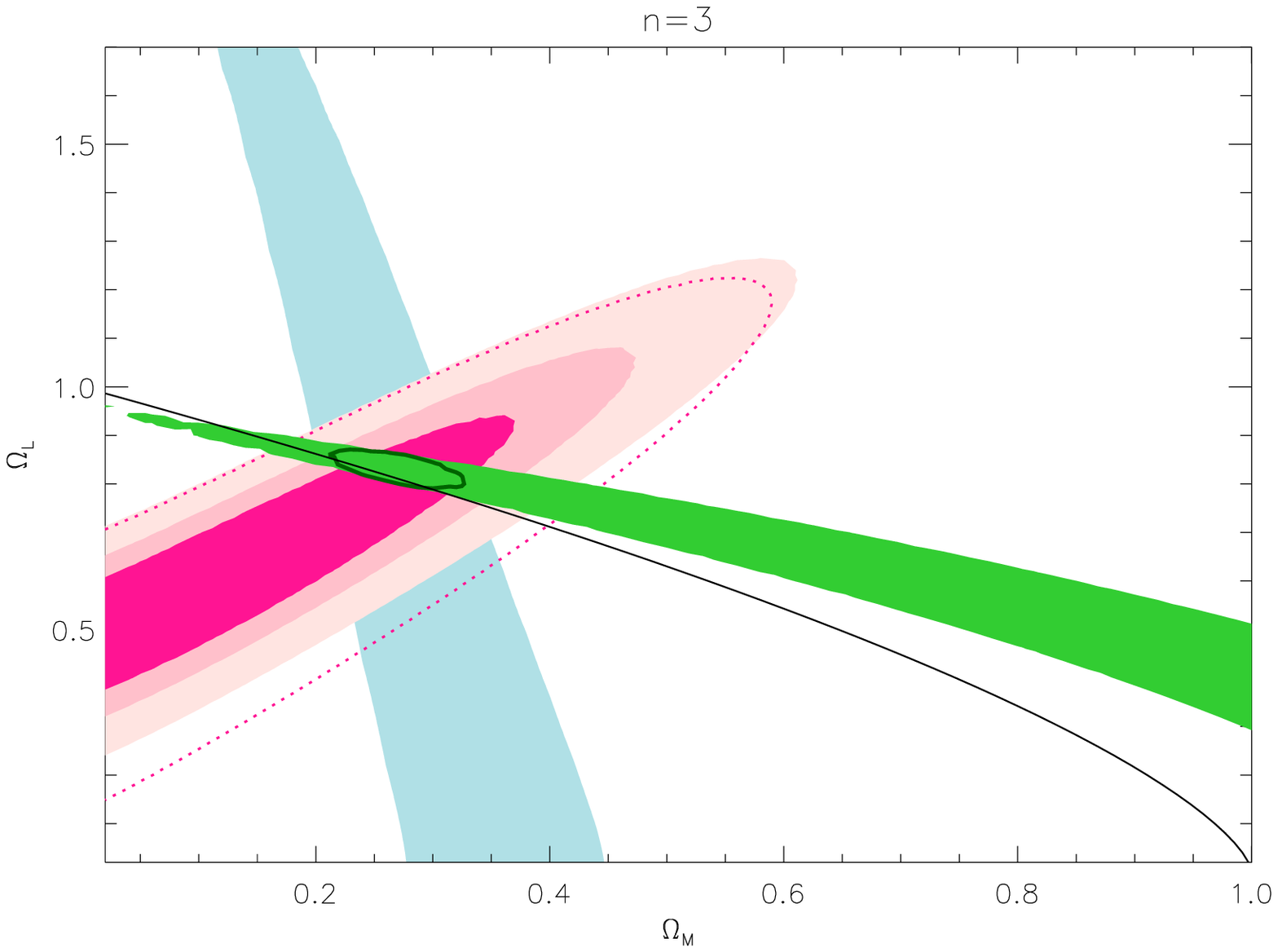}& 
\includegraphics[height=6cm,width=8cm]{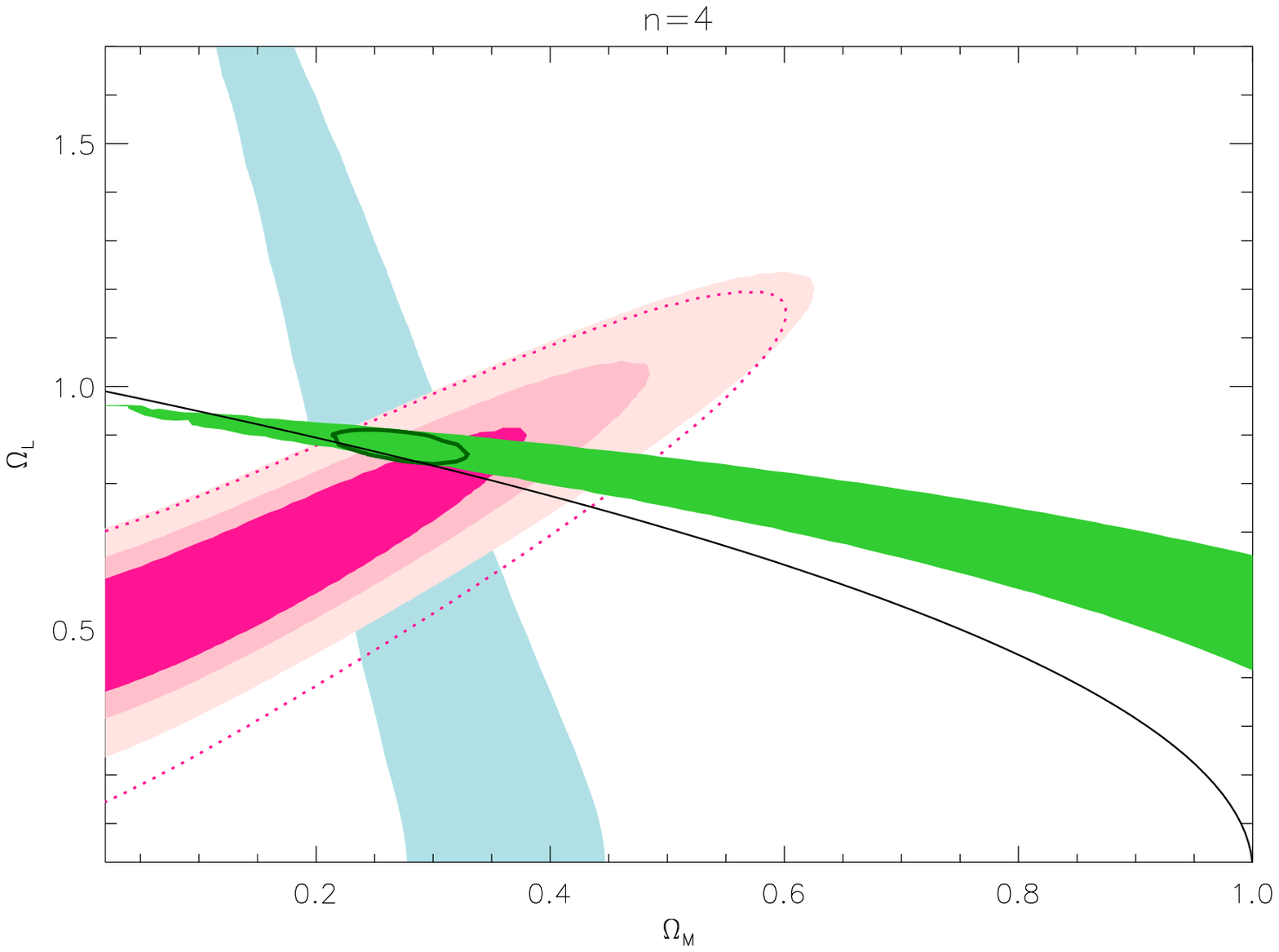}\cr
\end{tabular}    
\caption{Supernovae from ESSENCE, SNLS and nearby sample analysed with the MLCS2k2 method compared with models described by equation \ref{curvgeneral} with $\beta=0$ which include the basic DGP model ($n=1$).  Because $\Lambda$CDM is at this level identical to the $n=2$ case we plot the different models in order of increasing $n$.  The pink concentric ellipses correspond to the 68\%, 95\% and 99\% confidence regions from the latest supernova data (\ref{lumdim}) when the peculiar velocity error is not yet included, the blue bands border the 99\% confidence region for the baryon acoustic peak data (\ref{bao}) while the green region borders the 99\% confidence region for the CMB shift parameter (\ref{cmb}).  The pink dotted line corresponds to the 99\% confidence region for the supernova data when we have included the 400 km/s error in the redshift. The black line corresponds to spatially flat universes.
\label{MLCSdgp} }
\end{figure}

\begin{figure}
\begin{tabular}{cc}
\includegraphics[height=6cm,width=8cm]{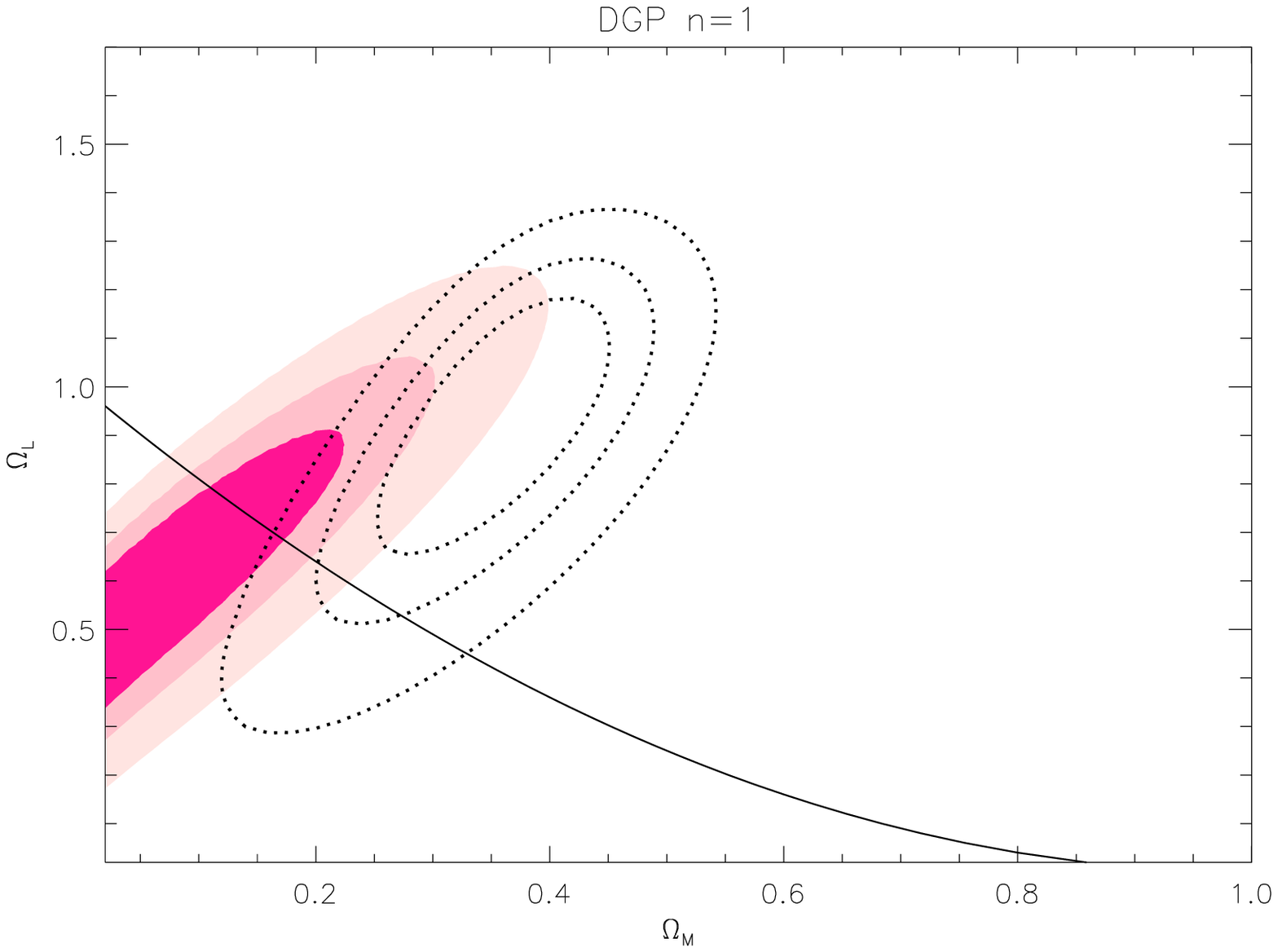}& 
\includegraphics[height=6cm,width=8cm]{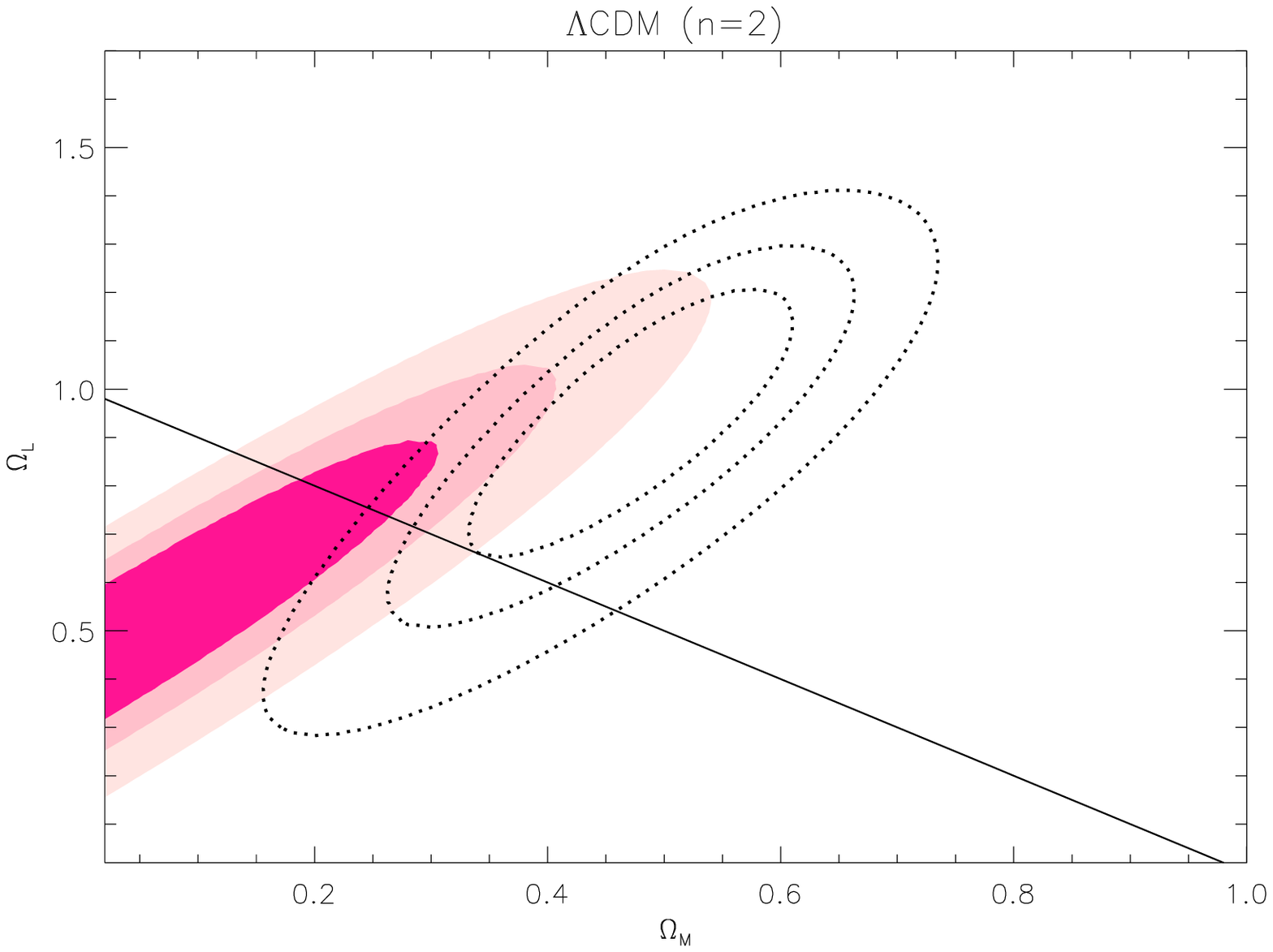}\cr
\end{tabular}    
\caption{Comparison between the results of fitting DGP and $\Lambda$CDM to the SNLS and ESSENCE supernova data set (filled in 68\%, 95\% and 99\% confidence regions) and the Riess 07 Gold set (dotted lines). The solid black line corresponds to spatially flat universes.
\label{Riess} }
\end{figure}

Figure \ref{MLCSdgp} shows the comparison of the data (supernovae,
galaxy survey baryon oscillation and CMB) with the DGP model and its
variants where the supernova data has been analysed using the MLCS2k2
method.  The confidence regions reflect the reported 
{\em statistical} uncertainties of the various measurements. The potential
impact from systematic effects is not addressed by this analysis.

Note that at this level of expansion history the $n=2$ DGP
variant is identical to $\Lambda$CDM although presumably perturbations
would grow very differently in the two models.  The actual $\chi^2$
values corresponding to the two different models DGP (n=1) and
$\Lambda$CDM (n=2), as well as the higher dimensional generalisations
for n=3 and n=4 are listed in the table \ref{chitable} and
\ref{chitablenew} . The effects of adding the velocity errors are of
course lower minimum $\chi^2$ values as well as a small shift in the
best fit parameter values, explaining the increased area of the
supernova confidence regions and their shift in parameter space seen
in figure \ref{MLCSdgp}.

\begin{table}
\begin{center}
\begin{tabular}{|l|c|c|c|c|c|}
\hline
$\beta=0$&\multicolumn{4}{|c|}{MLCS (162 data points)}\\
\cline{2-5}
 &n=1& $\Lambda$CDM (n=2) &n=3&n=4\\
\hline
$\chi^2_{min}$(SNe)& 188 & 188 & 188& 188\\
$\chi^2_{min}$(SNe+flat)& 188 & 188 & 189& 189 \\
$\chi^2_{min}$(SNe+flat+CMB) & 200 & 188 & 190& 197\\
$\chi^2_{min}$(SNe+flat+CMB+BAO) &215 & 192 & 191& 197 \\
\hline
\end{tabular}
\end{center}
\caption{Best fit $\chi^2$ values for the fits of different data sets to the various models.  SNe is the supernova data set (analysed using the MLCS2k2 procedure) without inclusion of the redshift error due to the peculiar velocities, next flatness is assumed then the CMB shift parameter is added to the data set.  Finally we include the Baryon acoustic oscillation result (BAO) from the Sloan Digital Sky Survey.}
\label{chitable}
 \end{table}

\begin{table}
\begin{center}
\begin{tabular}{|l|c|c|c|c|c|}
\hline
$\beta=0$&\multicolumn{4}{|c|}{MLCS (162 data points)}\\
\cline{2-5}
 &n=1& $\Lambda$CDM (n=2) &n=3&n=4\\
\hline
$\chi^2_{min}$(SNe)& 160 & 160 & 160 & 160\\
$\chi^2_{min}$(SNe+flat)& 161 & 161 & 161 & 162  \\
$\chi^2_{min}$(SNe+flat+CMB) & 171 & 161 & 163 & 170\\
$\chi^2_{min}$(SNe+flat+CMB+BAO) &180 & 163 & 164& 170 \\
\hline
\end{tabular}
\end{center}
\caption{Same as table \ref{chitable} except here the redshift error of 400 km/s has been included in the supernova data set.}
\label{chitablenew}
 \end{table}

 All models can fit the supernova data alone equally well, and more or less equally can fit the data assuming flatness.  The requirement of having to fit the CMB data singles out $\Lambda$CDM as being slightly favoured over the other models, although the $\chi^2$ values are such that the DGP model is roughly 1-2 $\sigma$ disfavoured, depending on whether the peculiar velocity error has been included or not, which is not statistically significant enough to allow us to claim that the model is ruled out.  
 
 If we add the baryon oscillation feature to the data we find that we do seem to be able to disfavour the DGP model at the 1.5-3 $\sigma$ level, again depending on what errors have been included, but as we have already stated in section \ref{difdata}, the reconstruction of the peak in the correlation function from redshift space depends upon assumptions which are not valid in cosmologies where the equation of state is varying over time.  At some point in the future it may be interesting to investigate such systematic effects quantitatively. 
 
Note also that even if we from the last row of table \ref{chitable} draw the conclusion that the DGP model is ruled out at a 3 $\sigma$ level, the $\Lambda$CDM model is then disfavoured at the 2 $\sigma$ level. Looking instead at table \ref{chitablenew} the $\chi^2$ values obtained for DGP and $\Lambda$CDM give goodness of fit of 16\% and 46\% respectively. 

Figure \ref{Riess} shows a comparison between the results of fitting
 the DGP model and the $\Lambda$CDM model respectively to the Riess 07
 gold set, used in \cite{scoop}, and the SNLS and ESSENCE data set,
 used in this paper. Clearly, there are inconsistencies between the 
 two published sets of SN data and we have therefore not attempted to 
 combine them.
We see that for supernova data and a flat prior
 only, the Riess data makes some distinction distinction 
between the two models while
 the SNLS data does not. It is on including the prior of the CMB shift
 parameter that the DGP and $\Lambda$CDM are equally well favoured by
 Riess data while the result in this paper is that DGP looks slightly
 less favoured than $\Lambda$CDM.
 
 \subsection{Best fit values of $n$.}
 
An interesting exercise is to take the case $\beta=0$ and allow the number of extra dimensions $n$ to be a free, non-integer parameter.  We then calculate which values of $n$ fit the data best when we allow $\Omega_M$ to be a free parameter.  

\begin{figure}
\begin{tabular}{cc}
\includegraphics[height=6cm,width=8cm]{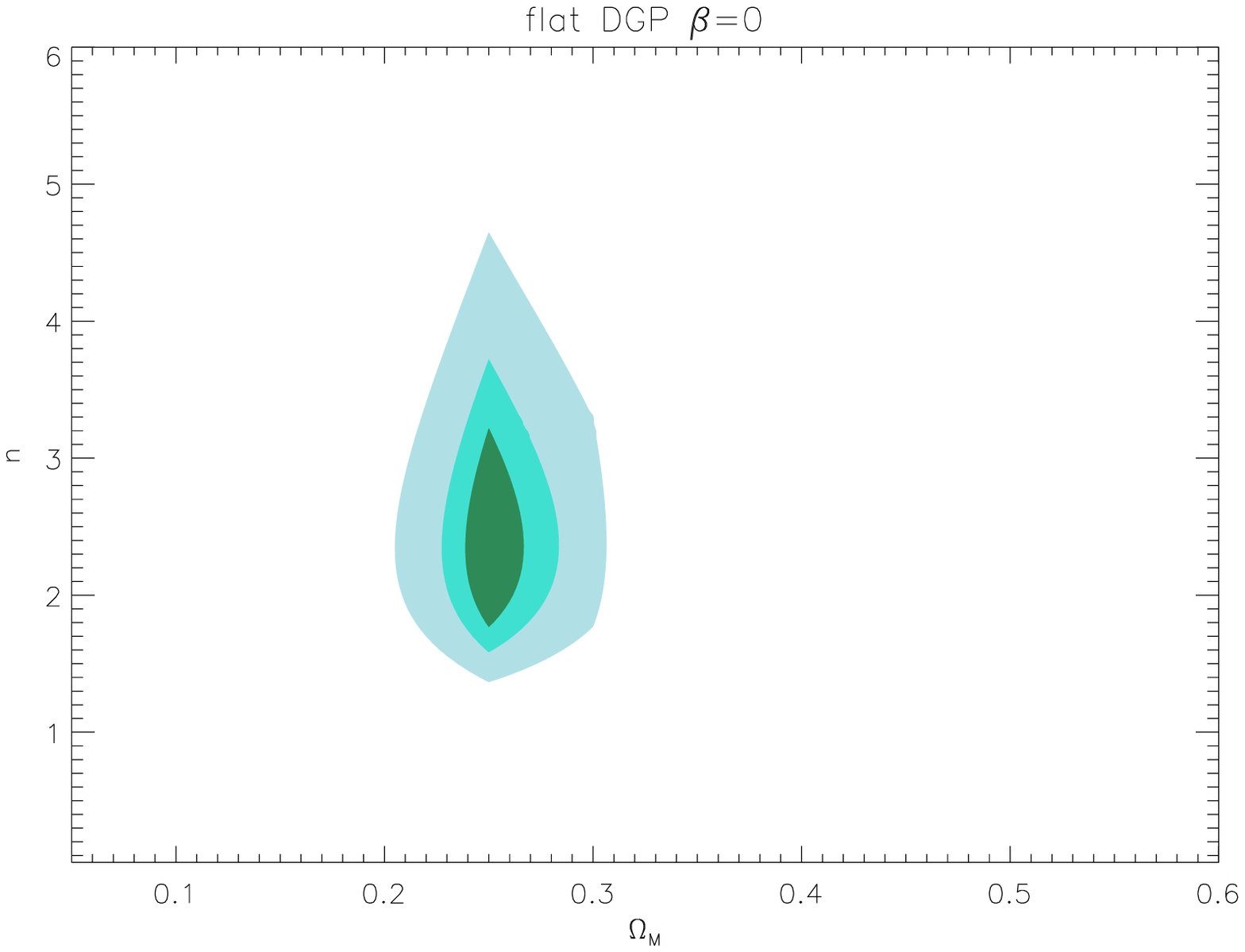}& 
\includegraphics[height=6cm,width=8cm]{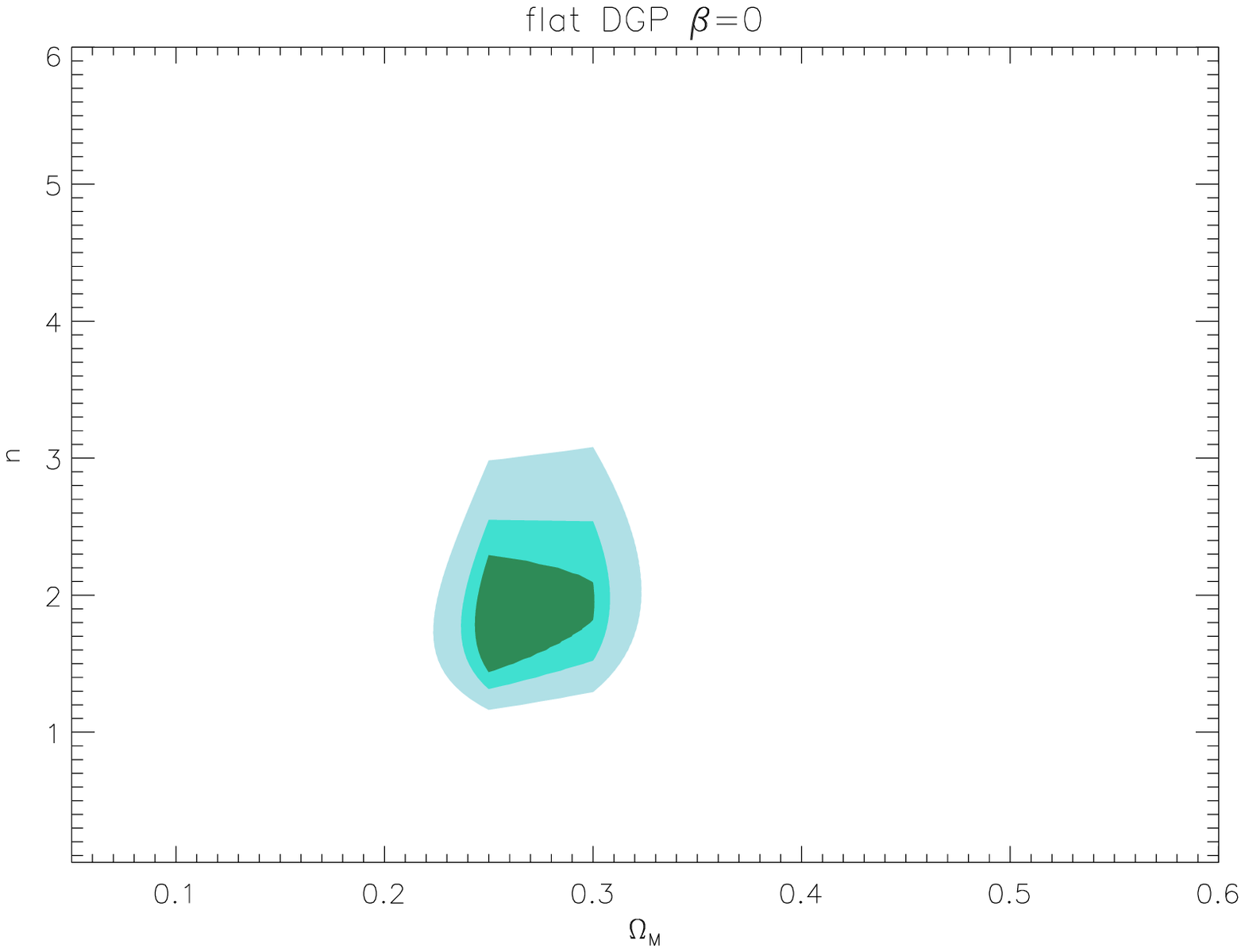}\cr
\end{tabular}    
\caption{Best fit values of $n$ for $\beta=0$ and different values of the matter density $\Omega_M$.  Flatness is assumed and the priors of CMB and BAO included. On the left the supernova data used is analysed by MLCS2k2 and on the right the data analysed by SALT.}
\label{bestn}
\end{figure}

Figure \ref{bestn} shows us that the best fit region when $\beta=0$ occurs for values of $n$ between 1.5 and 3.  The case $n=2$ corresponds exactly to a cosmological constant (whether $\beta=0$ or not), so this tells us that without $\beta$ the best fit to the data lies somewhere around the $\Lambda$CDM models, in agreement with the previous conclusions of \cite{fairbairngoobar} although the preferred value of $n$ may have increased slightly.

\subsection{Models with non-zero $\beta$}

As stated earlier, for the higher dimensional generalisations of DGP we should include the parameter $\beta$ which, while required for theoretical consistency, is an extra free parameter which
renders the theory less predictive and makes it easier to fit the
data.  For this reason, for the higher dimensional cases, we will
consider spatially flat universes with non zero $\beta$.  

The original DGP model corresponds to $n=1$ and in that case there is no need for a regularisation parameter $\beta$.  In terms of large scale geometry of space time the $n=2$ case is equivalent to $\Lambda$CDM with or without $\beta$.  We therefore only consider $n=3,4,5,6$.

In all of these four models, the confidence region in the parameter space of $\Omega_M$ and $\Omega_L$ centers around $\Omega_M=0.26$. Taking this as our value for $\Omega_M$, we obtain a minimum value of $\Omega_L$ for each $n$ by looking at the 95$\%$ confidence level away from the best fit, corresponding to a minimum value of $\beta$ that increases for increasing $n$, as shown in table \ref{betatable}.
\begin{table}
\begin{center}
\begin{tabular}{|c|c|c|c|c|}
\hline
95\% & $n=3$ & $n=4$ & $n=5$ & $n=6$\\
\hline
$\Omega_L$ $>$ & (0.8) & 0.95 & 1.15 & 1.25 \\
$\beta$ $>$ & 0 & 0.23 & 0.74 & 1.05 \\
\hline
\end{tabular}
\end{center}
\caption{Minimum values of $\Omega_L$ and $\beta$ (via equation (\ref{constraint}) with $\Omega_M=0.26$) with 95 $\%$
confidence.   For n=3, the minimum value of $\Omega_L$ is restricted only
through the cutoff where $\beta$ goes negative. \label{betatable}}
\end{table}

\begin{figure}
\begin{center}
\includegraphics[height=9cm,width=12cm]{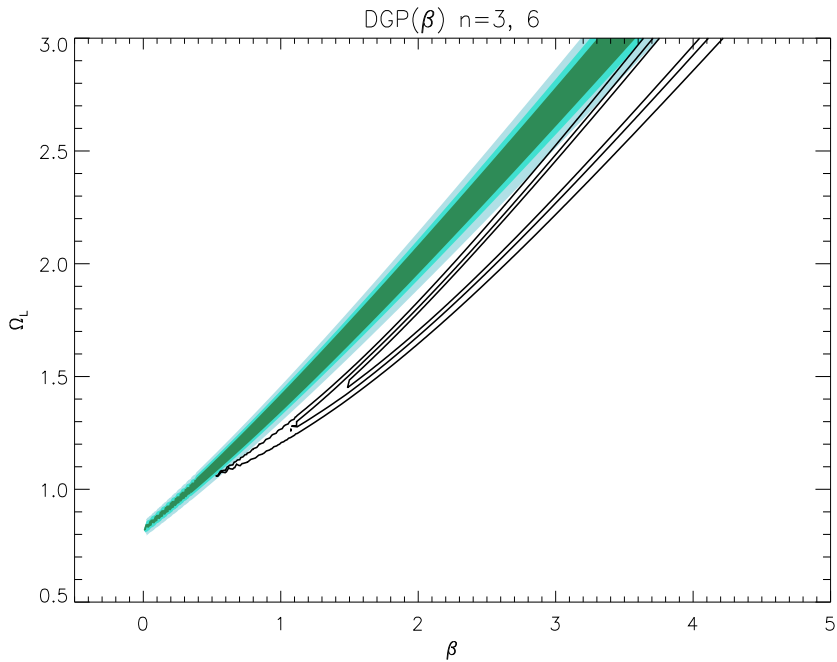}
\end{center}
\caption{\label{betafig} Constraints on the higher dimensional versions of DGP taking into account the regularisation parameter $\beta$. The filled in contours are the confidence levels for the combined $\chi^2$ for the $n=3$ case (see note at beginning of section).   The combined $\chi^2$ for the $n=6$ case is also plotted for comparison using black lines showing that for this case, zero $\beta$ is significantly disfavoured.}
\end{figure}

In figure \ref{betafig} we also show how the different cosmological constraints cut into the parameter space of $\beta$ and $\Omega_L$ by plotting the confidence regions for two of the higher dimensional models.

This shows that for a higher number of dimensions, the fit gets worse and one will have to increase the value of $\beta$ in order to get closer to the cosmological constant case which fits the data well. As $\beta$ becomes more dominant, the dark-energy term in eq. (\ref{dgpe})  approaches $\Omega_L/\beta$ and we can see from figure \ref{betafig} that the gradient approaches $\approx 0.7$ for larger $\beta$, which agrees with the usual best-fit value of $\Omega_{\Lambda}$.

\section{Summary and Conclusions}

Data from the Supernova legacy survey analysed with the SALT method has previously suggested that the DGP model is marginally disfavoured relative to $\Lambda$CDM \cite{fairbairngoobar}.  At the same time, a larger data set including recent data from the Hubble space telescope and analysed using the MLCS approach has led to the conclusion that the DGP model is in fact perfectly safe \cite{scoop}.

In this work, we have looked at the SNLS data and new data from the
ESSENCE collaboration analysed with the MLCS2k2 algorithm as reported
in \cite{essence}. 
Combination
of this data with the CMB constraint suggests that the DGP model is
slightly disfavoured, and becomes more dis-favoured if one
can treat the baryon acoustic peak as a valid data point.  The
magnitude of the change in the position of the acoustic peak in the
galaxy correlation function when one moves from a background cosmology
with a constant equation of state to a DGP universe depends both upon
the different geometries and the way that structure grows in those
universes.  Since the growth of structure in a DGP universe may be
rather different from in $\Lambda$CDM we are not able at this stage to
say whether or not the DGP model seems to be marginally or
significantly disfavoured using this data.  Either way, we find
that the tests of the DGP model yield significant differences when
using the ESSENCE supernova data and the ``gold set'' \cite{scoop,riess07}.

The fact that the two data sets lead to different conclusions about
the same model is very interesting and outlines the challenges which
need to be overcome in order to move into the era of precision Dark
Energy measurements.

\ack
We are grateful for discussions with Cedric Deffayet and Gia Dvali whilst doing this work.
MF thanks the Swedish Research Council and the Perimeter Institute for Theoretical Physics for their hospitality.
AG would like to acknowledge support by the Swedish Research Council and 
the G\"oran Gustafsson Foundation for
Research in Natural Sciences and Medicine.

\end{document}